\documentclass[structabstract]{aa}  
\usepackage{graphicx}
\usepackage{natbib}
\usepackage{txfonts}
\newcommand{\jb}{{\sl J}}
\newcommand{\hb}{{\sl H}}
\newcommand{\kb}{{\sl K}$_S$}
\newcommand{\mic}{$\mu$m}

\begin{document}
   \title{The extinction map of the OMC-1 molecular cloud behind the Orion Nebula}


   \author{G. Scandariato\inst{1}
          \and
          M. Robberto\inst{2}
          \and
          I. Pagano\inst{3}
          \and
          L. A. Hillenbrand\inst{4}
          }

   \institute{Dipartimento di Fisica e Astronomia, Universit\`a di Catania, Italy\\
              \email{gas@oact.inaf.it}
         \and
             Space Telescope Science Institute, Baltimore, MD 21218\\
         \and
          	INAF Osservatorio Astrofisico di Catania, via S.\ Sofia 78, 95123 Catania, Italy\\
		  \and
		  	Astrophysics, California Institute of Technology, Pasadena, CA91125
             }


 
  \abstract
{The Orion Nebula and its associated young stellar cluster are located at the front-side of the optically thick \object{OMC-1} molecular cloud. In order to disentangle the cluster members from background contamination, it is important to know the extinction provided by the OMC-1, which is poorly known, the available measurements yielding contradictory results.}
{Our main goal is to derive a new extinction map of the OMC-1, obtaining information about the structure of the OMC-1 and the \object{Orion Nebula Cluster}.}
{The most recent near-infrared catalog of stars is used to study the distribution of reddening across a ~0.3 $\rm deg^2$ area covering the Orion Nebula Cluster. On the basis of the observed $(H,H-K_S)$ diagram, we establish a criterion for disentangling contaminants from bona-fide cluster members. For contaminant stars, interstellar reddenings are estimated by comparison with a synthetic galactic model. A statistical analysis is then performed to consistently account for local extinction, reddening and star-counts analysis.}
{We derive the extinction map of the OMC-1 with angular resolution $<$5$\arcmin$. We also assemble a sample of candidate cluster members, for which we measure the extinction provided by the nebular environment. These extinction measurements are analyzed similarly to the contaminant sample, and an extinction map of the \object{Orion Nebula} is derived.}
{The extinction provided by the OMC-1 is variable on spatial scales of a few arcminutes, while showing a general increase from the outskirts ($A_V\sim6$) to the direction of the \object{Trapezium} asterism ($A_V\gtrsim30$). The Orion Nebula extinction map is more irregular and optically thinner, with $A_V$ of the order of a few magnitudes. Both maps are consistent with the optical morphology, in particular the Dark Bay to the north-east of the Trapezium. Both maps also show the presence of a north-south high-density ridge, which confirms the filamentary structure of the Orion molecular complex inside which star formation is still taking place.}


   \maketitle
%

\section{Introduction}
The closest event of massive star formation is occurring at present in the direction of the Galactic anticenter ($l=209^\circ, b=-19^\circ$), in the Orion constellation. The Orion Nebula (ON, M42) represents the most spectacular signature of star formation activity in this region. The ON is a blister HII region carved into the OMC-1 giant molecular cloud by the UV flux emitted by a handful of OB stars, the so-called Orion \lq\lq Trapezium\rq\rq\ \citep{Mue08, odell}. The Trapezium stars are the most massive members of a rich ($n\simeq 2000$ members, \citet{Mue02}) cluster of young (1-3~Myr old, \cite{DaRio10}) Pre-Main-Sequence objects (Orion Nebula Cluster, ONC). Given its youth, vicinity, and low foreground extinction, the observed luminosity function of the ONC can be converted into a true initial mass function with relatively modest assumptions \citep[and references therein]{Mue02}. It is largely for this reason that the Orion Nebula and its associated cluster are regarded as a critical benchmark  for our understanding of the star formation process.
 
In order to build a reliable luminosity function, especially in the substellar regime (brown dwarfs and planetary mass objects) and to explore its spatial variations with the distance from the cluster center, it is necessary to remove the contribution of non-cluster sources. In principle, this requires the acquisition of thousand of spectra of faint sources distributed over the bright nebular background. On the other hand, the number of contaminant sources, both galactic and extragalactic, can be estimated using the most recent models for stellar and galaxy counts at various wavelengths. The main complication in this case arises from the presence of the OMC-1, which provides a backdrop to the ONC of high and non-uniform extinction. Deriving an accurate extinction map would be beneficial not only to better discriminate the ONC membership, but also to understand the 3-D distribution of the cluster, still partially embedded within the ONC, and the evolutionary history of the region.

In this paper we present a reconstruction of the OMC-1 extinction map based on the analysis of the recent near-IR (NIR) source catalog of \citet{paperI} (R10 hereafter). In Section~\ref{sec:overview} we briefly review the previous studies relevant to the Orion Nebula region. In Section~\ref{sec:avmap} we illustrate our statistical method to disentangle background stars from the cluster population. By combining an estimate of the interstellar extinction affecting each contaminant star with the source count density, we derive the OMC-1 extinction map. In Sect.~\ref{sec:foreground} we apply a similar statistical procedure to the candidate cluster members, deriving an extinction map for the dust in the foreground Orion Nebula. Finally, in Sect.~\ref{sec:discussion} we compare our maps to previous studies, and we briefly discuss their main features and similarities to stellar distributions.


\section{Overview of previous studies}\label{sec:overview}

A number of previous studies provide results relevant to the issue of the galactic reddening in the direction of OMC-1.

\citet{Sch98} (SFD98 hereafter) combined the \textit{COBE}/DIRBE observations (100~\mic\ and 240~\mic) and the \textit{IRAS}/ISSA observations (100~\mic) to obtain a full-sky 100~\mic\ map with $\sim6\arcmin$ resolution. On the basis of the correlation between the Mg line strength and the $(B-V)$ color of elliptical galaxies, they were able to calibrate their column-density map to a $E(B-V)$ color excess map. Their all-sky reddening map, available through the NASA/IPAC Infrared Science Archive\footnote{http://irsa.ipac.caltech.edu/applications/DUST/} Dust Extinction Service web page, represents  the benchmark for the following studies of our region.

The accuracy of the SFD98 maps has been analyzed by \cite{Arce99}, who compared the extinction map of the Taurus dark cloud with the extinction maps obtained using four other methods, i.e.\ 1) the color excess of background stars with known spectral types; 2) the ISSA 60 and 100~\mic\ images; 3) star counts; and 4) optical color excess analysis. These four methods give  similar results in regions with $A_V\leq4$. Their comparison shows that the extinction map derived by SFD98 tends to overestimate the extinction by a factor of $1.3\div1.5$. They ascribe this discrepancy to the the calibration sample of ellipticals used by SFD98, which  containing few sources with $A_V>0.5$ may lead to a lower estimate of the opacity. \cite{Arce99} also find that when the extinctions shows high gradient ($\gtrsim10^m/deg$, see their Fig.\ 1), its value is generally smaller than that given from SFD98, and argue that the effective angular resolution of the SFD98 map could be somewhat larger than 6$\arcmin$. 

\cite{dobashi05} obtained another all-sky absolute absorption map applying the traditional star-count technique to the  "Digitized Sky Survey I" (DSSI) optical database, which provides star densities in the range $\sim$1--30 arcmin$^{-2}$. Comparing the SFD98 maps to their results in the Taurus, Chameleon, Orion and Ophiucus complexes, they found that SFD98  extinction values are generally up to 2-3 times larger (see their Figs. 46, 47 and 48). A least-squares fit of the two extinction estimates in the $A_V\leq4$ gives a proportionality coefficient of 2.17, while the same analysis in the $A_V\geq4$ suggests a coefficient of $\sim3$. The authors discuss the possible causes of discrepancy. First, they stress that the SFD98 map are mostly  sensitive to the total dust along the line of sight, while their maps measure the extinction due to the nearby dust, as the optical thickness is much larger in the visible than in the far infrared. Second, they point out that the low resolution ($\sim1\deg$) of the temperature map adopted by SFD98 cannot reproduce the typical high spatial frequency temperature variations across dark dense clouds. Finally, they suggest that SFD98 do not account for enhancement in the far infrared emission by fluffy aggregates: this excess emission could explain the inconsistency between SFD98 map and maps derived using other methods. Unfortunately, DSSI optical plates saturate in proximity of the Trapezium cluster, so a large fraction of the ONC field is excluded from their analysis.

An extinction map limited to the inner $\sim$5\arcmin$\times$5\arcmin\ region of the ONC was provided by \citet{HC00} on the basis of the C$^{18}$O column density data of \citet{gold97}, with a 50\arcsec spatial resolution. A comparison of the extinction map they derive with the extinction obtained by SFD98 shows that that the SFD98 extinction is generally $\sim3$ times larger.

In summary, the SFD98 extinction map for the OMC-1 is still the only one covering the entire ONC field. However, its spatial resolution (6$\arcmin$) is limited and the accuracy, in a region as complex as the OMC-1, remains questionable.

In Sect.~\ref{sec:avmap} we overcome most of the issues listed above using photometric data in the NIR bands, as they offer several well known advantages. First, 90\% of the galactic population is made up of M dwarfs which, by virtue of their SED, radiate mostly in the NIR. Second, the extinction is lower at IR  wavelengths and therefore stars here can be more easily detected through optically thick clouds. Third, the high surface density of field stars allows to overcome the limited spatial resolution typical of far-infrared wide-field surveys used to measure diffuse dust emission. As a caveat, however, we anticipate that in our the number of detected background stars will  be limited by the brightness of the background, wich affects the completeness limit of the NIR survey (Sect.\ \ref{sec:avmap}). This effect will be taken into account in our analysis.

\section{The OMC-1 extinction map}\label{sec:avmap}

To compute a new extinction map of the OMC-1 region, we use the near-infrared photometry obtained by R10 at the CTIO/Blanco 4~m telescope in Cerro Tololo with the ISPI imager. The assembled photometric catalog contains \jb\hb\kb\ photometry in the 2MASS system for 7759 sources, 6630 labeled as point-like sources, spread over an area of $\sim$30\arcmin$\times$40\arcmin\ roughly centered on $\theta^1$Ori-C (RA=$05^h 35^m16.46^s$, DEC=$-05^{\circ} 23\arcmin23.2\arcsec$). While this catalog is slightly shallower than previous studies of the cluster core (e.g.\ \citet{HC00} or \citet{Mue02}), it covers a sky area $\sim$50 times larger, roughly centered on the Trapezium stars. We concentrate our analysis on the (\hb,\hb-\kb) color-magnitude diagram (CMD) of the $\sim$6000 point-like sources with both \hb\ and \kb\ measured photometry in the R10 catalog. \citet{Lombardi2001} show that the inclusion of the \jb\ magnitudes can reduce the noise of the extinction measurements by a factor of two, especially in regions characterized by low extinctions. Unfortunately, this is not our case. By using the full sample of stars detected in the \hb\ and \kb\ bands, we greatly increase the stellar density and therefore maximize the angular resolution of our map, since the R10 photometric catalog is shallower in the \jb-band. The R10 survey also lacks deep \jb-band observations of the North-East corner of their field, which makes the deep field covered by the the \jb-band smaller by about 9\% than the field covered in \hb\ and \kb-bands. Later in this paper we will include the \jb-band data, when we will consider the extinction map toward the ONC, which lays in the foreground of the OMC-1.

\begin{figure}
 \centering
 \includegraphics[width=.9\linewidth]{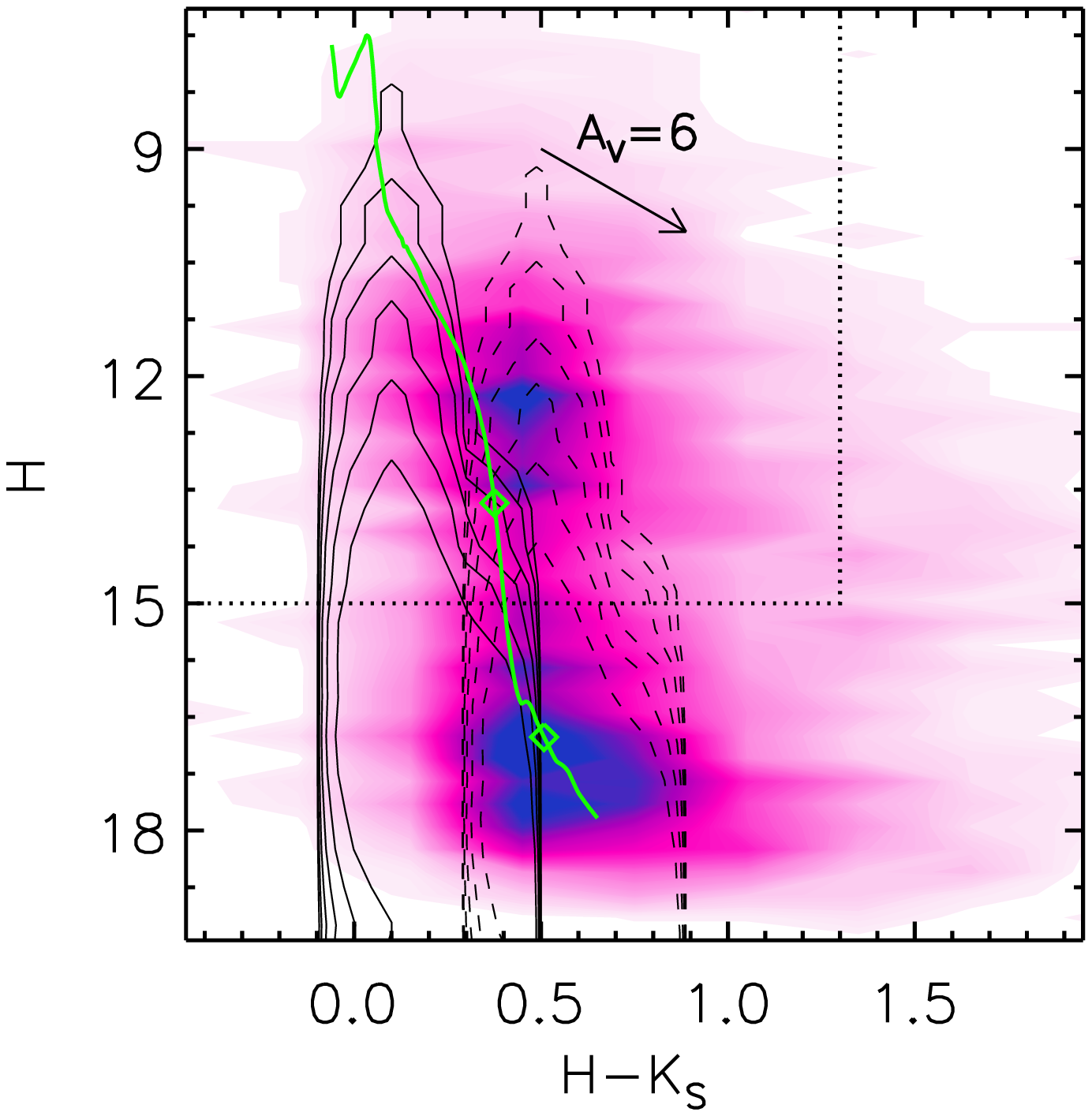}
 \caption{Density map in the (\hb,\hb-\kb) CMD of all point-like sources in the R10 catalog with measured \hb\kb\ photometry (color-filled areas), together with the 2~Myr isochrone computed by \citet{paperIMF} (solid green line, the diamond symbols correspond to 1M$_\odot$ and 0.1M$_\odot$ from top to bottom respectively). Two main populations are apparent: the ONC clustered at \hb=12 and the bona-fide background stars at \hb$\sim$17. The latter population is to be compared with the synthetic galactic population provided by \citet{Robin03} (solid contours) after accounting for the extinction provided by the OMC-1. The case of a constant $A_V$=6 across the cloud is shown in dashed contours, the highest ones roughly overlapping with the observed background population. The dotted line represents our safe selection criterion: any source fainter than \hb=15 or redder than \hb-\kb=1.3 is considered a field star, provided that its colors are not compatible with the reddened Meyer's locus (Sect.\ \ref{sec:foreground}).}\label{fig:contcm}
\end{figure}

The (\hb,\hb-\kb) CMD in Fig.~\ref{fig:contcm} shows a characteristic bimodal distribution: a first group of stars (the ONC) is clustered at \hb-\kb=0.5 and \hb=12, whereas a second group of fainter objects appears clustered around \hb$\sim$17. This second group also has a peak, but this is a selection effect, as the number of faint sources drops at $\hb\ga17.5$ in correspondence of the sensitivity limit of the R10 survey.  

Figure~\ref{fig:contcm} also shows a 2~Myr isochrone appropriate for the ONC (solid green line) and the density contour of the galactic stellar population along the line of sight of the ONC (solid contours). The galactic population has been derived from the Besan\c{c}on galactic model \citep{Robin03}\footnote{http://bison.obs-besancon.fr/modele/}, computed at the galactic coordinates of the ONC over the R10 survey area of 0.329 $deg^2$. For each star we computed synthetic photometry in the 2MASS system using the grid of atmosphere models of \citet{Allard2010}. Since the galactic model also provides for each star the distance and interstellar extinction, both parameters were accounted for in the synthetic photometry. 

Of the 4868 sources with \hb$\le$18 returned by the model, 4616 ($\approx$95\%) lie at distances larger than 420 pc. Their colors will generally be further reddened by the OMC-1. In Fig.~\ref{fig:contcm} we plot (dashed lines) the galactic model reddened by $A_V=6$, an \textit{ad-hoc} value chosen to shift the contours in the same region occupied by the second peak  at \hb$\sim$17 in the (\hb,\hb-\kb) diagram. As anticipated by R10, this confirms that the second peak in Fig.~\ref{fig:contcm} is fully compatible with the reddened galactic population. 

The peculiar morphology of the CMD, characterized by two well separated peaks associated to the ONC and the reddened galactic populations, allows us to assume that sources fainter than \hb=15 or redder than \hb-\kb=1.3 (respectively below and to the right of the dotted lines in Fig.~\ref{fig:contcm}) are largely background field star, being either too faint or too red to belong to the ONC, and vice versa for the other sources within the area. This is a first order criterion that can be further refined. Firstly, we can take advantage of the finer angular resolution of the ACS images obtained for the HST Treasury Program to discard 73 double stars  and 166 extended objects, either background galaxies or proplyds \citep{Ricci2008}. Moreover, by looking at the \jb\hb\kb\ color-color diagram for the subsample of stars with available \jb\ magnitude, we find 61 \lq\lq field\rq\rq\ stars well compatible with the locus of reddened circumstellar disks (Section \ref{sec:foreground}). These sources have been therefore moved to the cluster sample. Our final samples are thus made up of 1716 cluster sources and 1913 background galactic stars.

\begin{figure*}
\centering
\includegraphics[width=.4\linewidth,viewport=1 40 857 1100,clip]{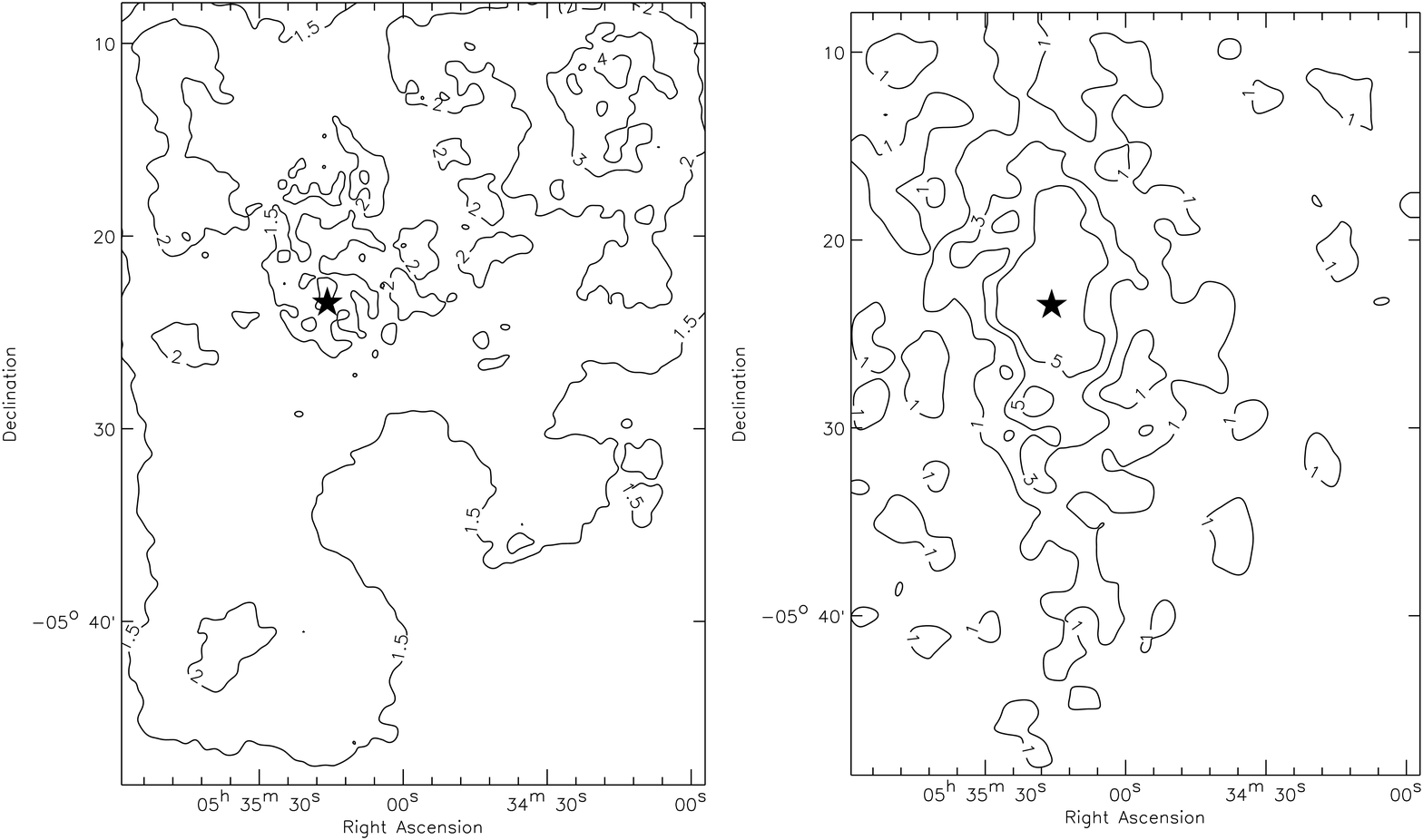}\hspace{1cm}
\includegraphics[width=.4\linewidth,viewport=874 40 1730 1100,clip]{densities.eps}
\caption{Projected surface density of bona-fide background stars (left panel) and of bona-fide cluster members (right panel), as selected in Sect.~\ref{sec:avmap}. The contour labels give the density of stars in $arcmin^{-2}$. In the direction of the Trapezium cluster (the star symbol indicates the position of $\Theta^1OriC$, assumed to be the cluster center) we find the minimum density of background stars, while it increases towards the outskirts of the cloud. This is indicative of the decreasing optical thickness of the OMC-1 with increasing projected distance from the cluster center. Conversely, the surface density of cluster members steeply increases with decreasing distance from $\Theta^1OriC$.}\label{fig:densities}
\end{figure*}

Figure \ref{fig:densities} shows the projected density contours for the two samples of background and cluster stars. The relatively low surface density of background stars in the direction of the Trapezium cluster indicates that the optical depth of the OMC-1 reaches its maximum behind the Trapezium. Vice versa, the density of bona-fide cluster members reaches its maximum in the direction of the Trapezium cluster and decreases towards the outskirts of the Orion Nebula, as already suggested by e.g. \citet{HillenbrandHartmann1998}. The density of stars, either background stars or cluster members, is critical in our statistical derivation of the extinction map of the OMC-1 (Sect.\ \ref{sec:computation}) and of the Orion Nebula (Sect.\ \ref{sec:foreground}) and drives the angular resolution of our derived maps.

It is clear that the two samples are not pure, e.g. there are brown dwarfs and/or planetary objects belonging to the cluster that may have been improperly included in the background sample. According to the previous studies of the ONC initial mass function \citep[see][and references therein]{Mue08}, the fraction of sources having substellar masses decreases with decreasing mass and therefore one may argue that the background sample is poorly affected by the presence of ONC member with very low masses. By constraining the stellar density to match the average extinction, we will further reduce the cases of spurious identifications. In any case, contamination will remain a source of uncertainty.

While it is possible to assume that the intrinsic color of a cluster star (neglecting disk excess emission) roughly 
corresponds to that of a source on the 2~Myr isochrone, for the background population there is no such isochrone. In the following sections we describe the statistical approach (Sect.~\ref{sec:single}) we have used to derive the extinction affecting each background star and consistently derive the extinction map of the OMC-1.

\subsection{The extinction affecting each background star}\label{sec:single}

\begin{figure*}
 \centering
 \includegraphics[width=.6\linewidth]{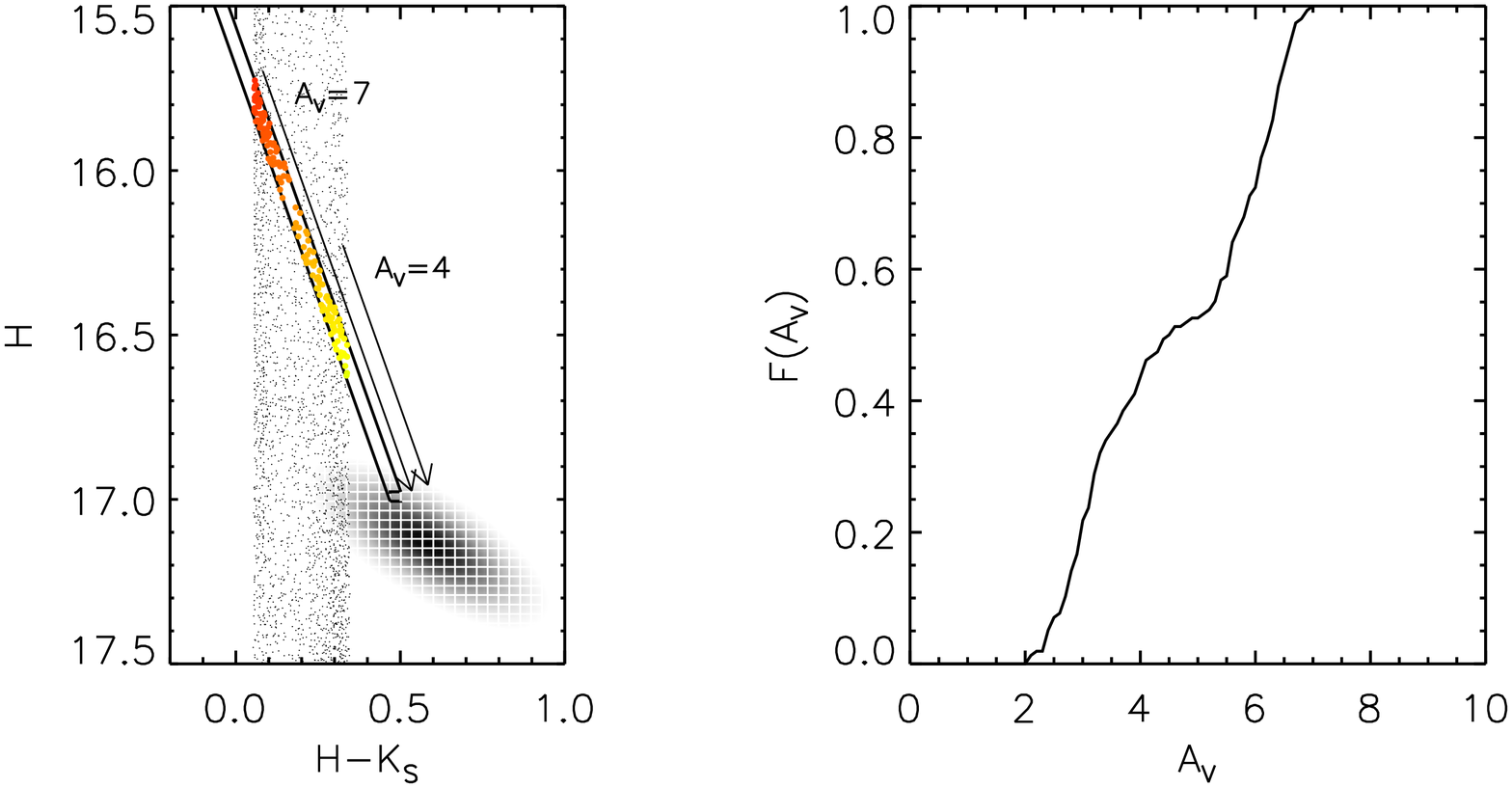}
 \includegraphics[width=.3\linewidth]{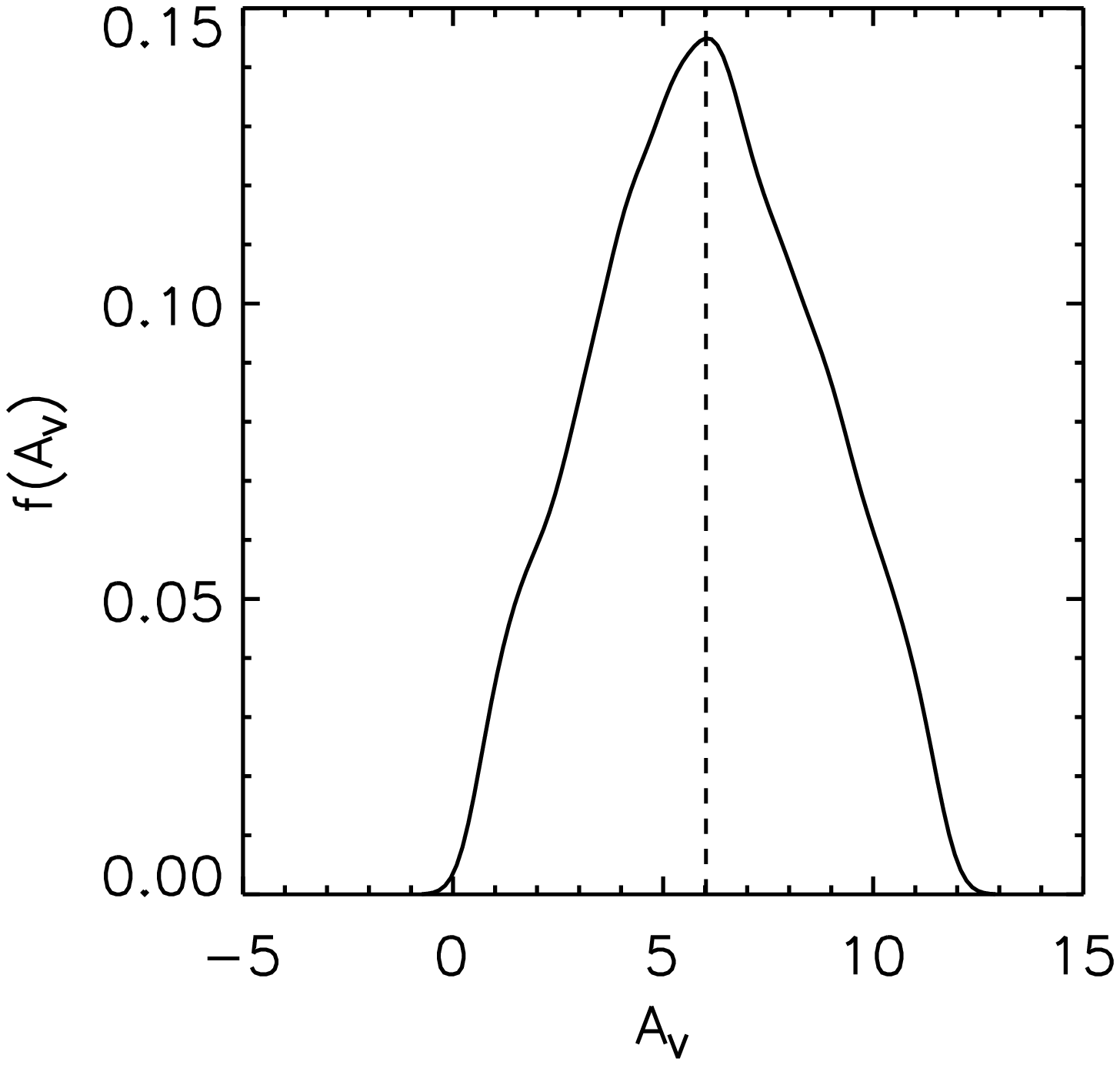}
 \caption{\textit{Left panel -- }Example of the dereddening algorithm for the star with ID=578 in the R10 catalog. The star is represented in the CMD by a regularly gridded 2D gaussian density distribution, and to each grid element it corresponds a stripe of the galactic model (represented in dots) parallel to the reddening direction. Dereddened candidates for the outlined grid point are coded with colors from yellow to red according to increasing $A_V$. \textit{Central panel -- }Cumulative distribution $F(A_V)$ for the colored synthetic galactic stars in the left panel. For each $A_V$, this function represents the fraction of candidate dereddened photometries with extinction $\leq A_V$. \textit{Right panel -- } Weighted probability density function $f(A_V)$ of the 40000 probabilistic $A_V$ estimates for the representative case shown in the left panel, computed using a gaussian kernel. The weighted mean of the sample (dashed vertical line) is our best estimate of the true extinction affecting the star.}\label{fig:avstripe}
\end{figure*}

 The basic idea of our method, similar to the one outlined by \citet{HC00}, is illustrated in  the left panel of Fig.~\ref{fig:avstripe}. Each star is represented in the CMD by a 2D probability distribution (the ellipse at the bottom right of the figure), where the shape of the distribution is due to the correlation between the \hb\ and \hb-\kb\ measures, with their uncertainties. This density distribution has been uniformly binned in a 20$\times$20 grid, resulting in 400 grid points.

We project each grid point backward in the reddening direction, tracing a stripe which intersects the galactic population model. Each grid point is thus associated to a certain number of galactic stars with different extinction values, without any restriction on the maximum extinction allowed. Each of these stars is a candidate to represent the dereddened photometry of the grid point. The central panel of Fig.~\ref{fig:avstripe} shows the cumulative distribution $F(A_V)$ of the reddening distribution of the candidates associated to the grid point shown in the left panel.

Using the cumulative distribution of $A_V$, for each grid point we randomly draw 100 $A_V$--values and assign to the full set of 100 estimates of $A_V$ a weight given by the 2D probability density distribution of the given grid point. Repeating this process for all grid points, we obtain a total of 400$\times$100=40000 probability estimates, each one with its own weight: the weighted mean of this sample gives the most likely estimate of the extinction affecting a given observed star. The right panel in Fig.~\ref{fig:avstripe} shows the weighted probability function $f(A_V)$ corresponding to the 40000 random estimates of $A_V$; to avoid data binning and for the sake of smoothness, the function is computed using a gaussian kernel density estimator \citep{Silverman1986}. The computation is repeated for all background sources, providing an extinction estimate, with relative uncertainty, for each star of the sample.

\subsection{Computation of the extinction map}\label{sec:computation}

To compute the OMC-1 extinction map, we project over  the original survey area of R10 a grid of points every  $100\times100$ pixels (corresponding to $30\arcsec\times30\arcsec$). For each grid point we select the 20 closest background stars within 1,000 pixels (5\arcmin). This limit is meant to prevent that, especially in regions with high extinction ($A_V>$15) where the density of background stars can be extremely low, the algorithm collects sources over an area too widespread. Thus, by setting an upper limit to the distance of background stars, we tend to preserve the angular resolution of the extinction map at the price of working with smaller statistical samples and therefore with locally more uncertain results. The value of 5\arcmin\ has been tuned to make sure than even in the regions with the highest extinction and lowest background density the algorithm will find a few background stars within that distance. This limit is also comparable to the scale size of major nebular structures seen in the NIR images.

Each sample of 20 (or less) background stars is averaged using an iterative sigma-clipping routine, deriving a robust estimate of both the mean local extinction $\bar{A_V}$ and standard deviation $\sigma_{\bar{A_V}}$. 
This would represent the solution to our problem, except that we want to make sure that the observed surface density and brightness distribution of the background sample is still compatible with the model. To this purpose, we apply the $\bar{A_V}$ extinction to the synthetic galactic model and compare the local density of the reddened model population to the observed one. 

The predicted density of the reddened galactic population has to be adjusted for the completeness of the survey, which statistically accounts for the success/failure of detection based on the magnitude of both stars and local background. In R10 we computed the completeness levels in three concentric regions centered on $\theta^1$Ori-C, neglecting any variation inside each region. For this study we improve the completeness estimates using a locally-computed completeness level, as described in Appendix \ref{sec:completeness}: for each grid point in the extinction map, our new approach allows us to compute the appropriate completeness adjustment.

\begin{figure*}
\centering
 \includegraphics[width=.35\linewidth]{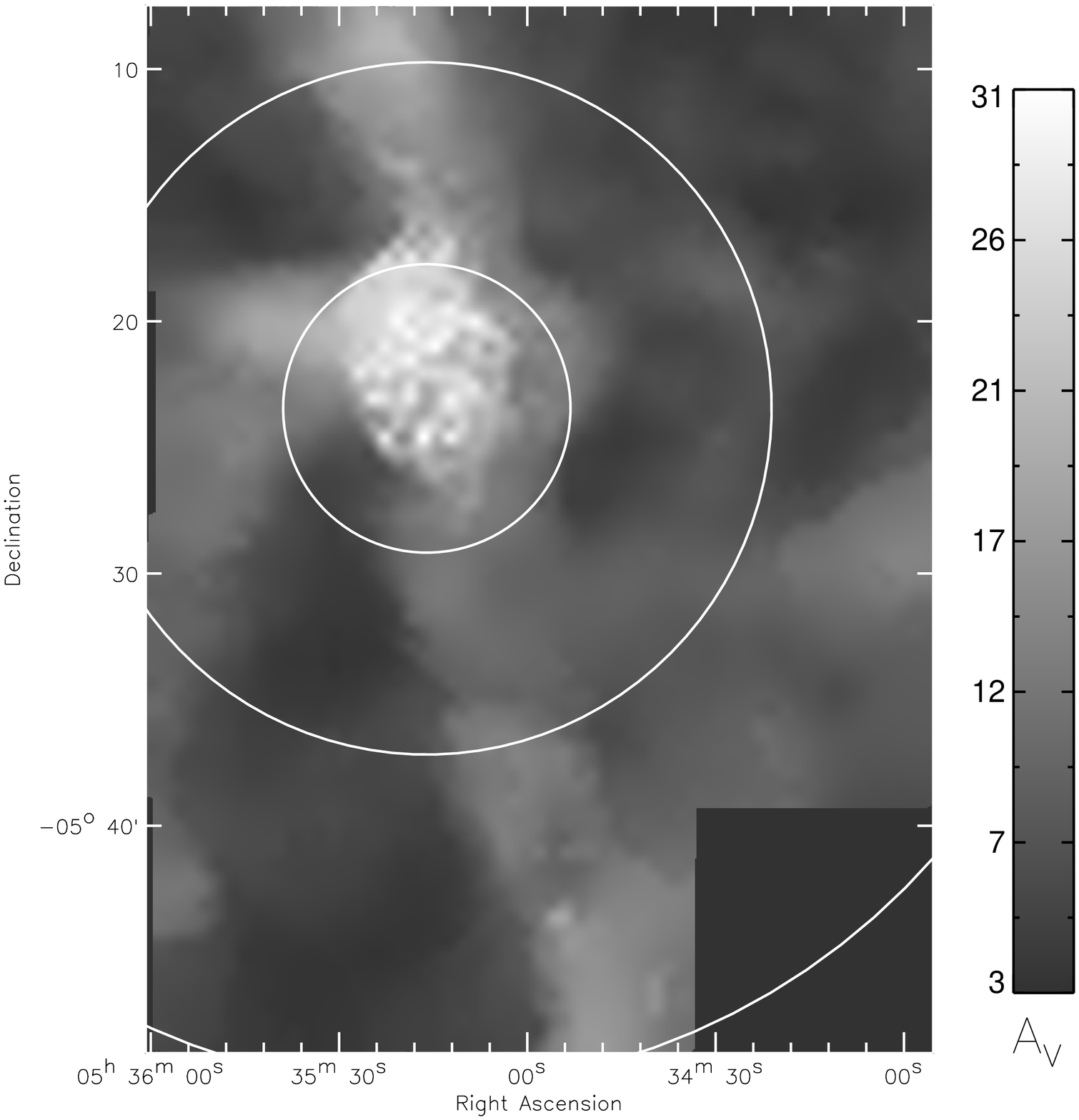}
 \includegraphics[width=.29\linewidth]{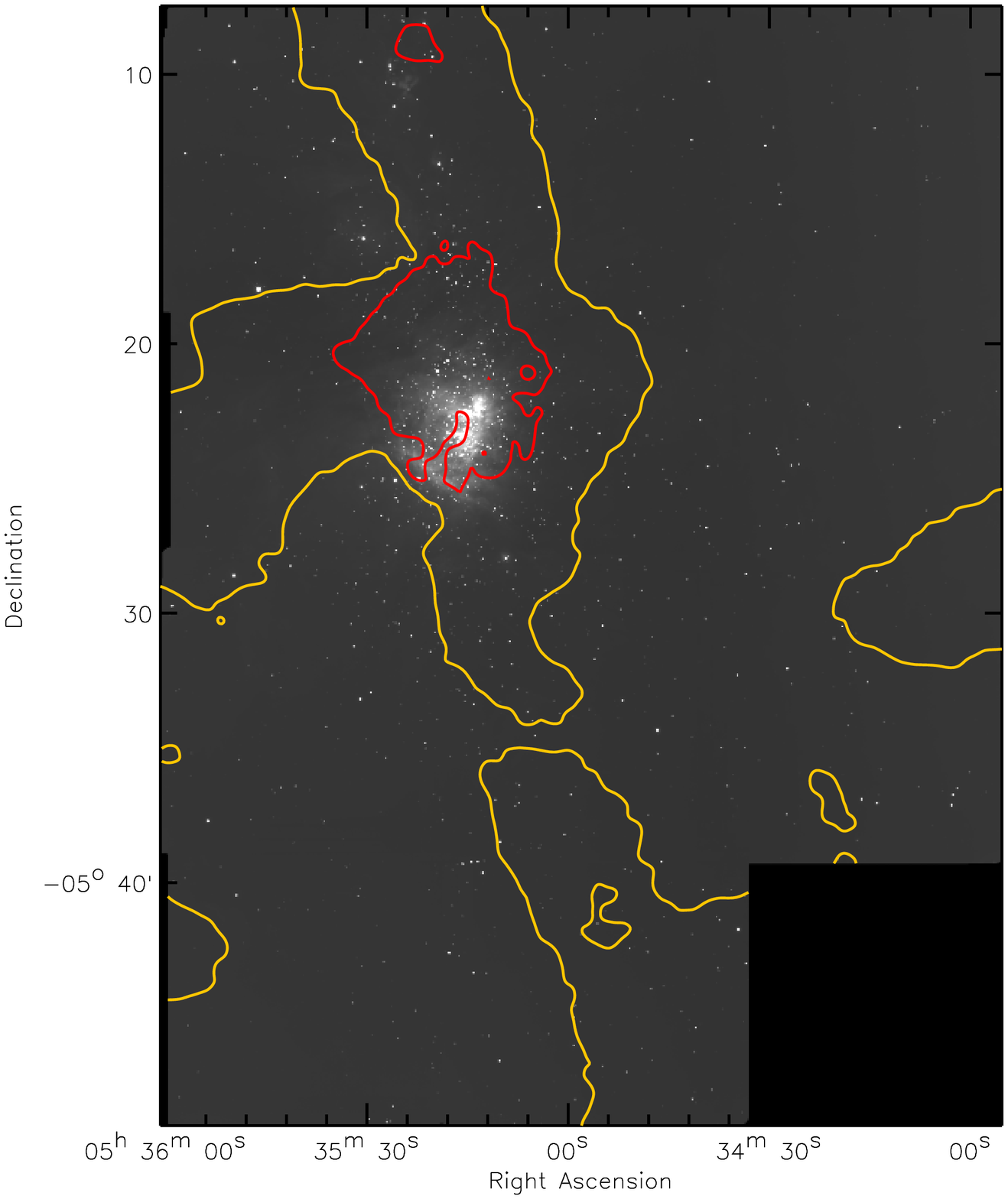}
 \includegraphics[width=.35\linewidth]{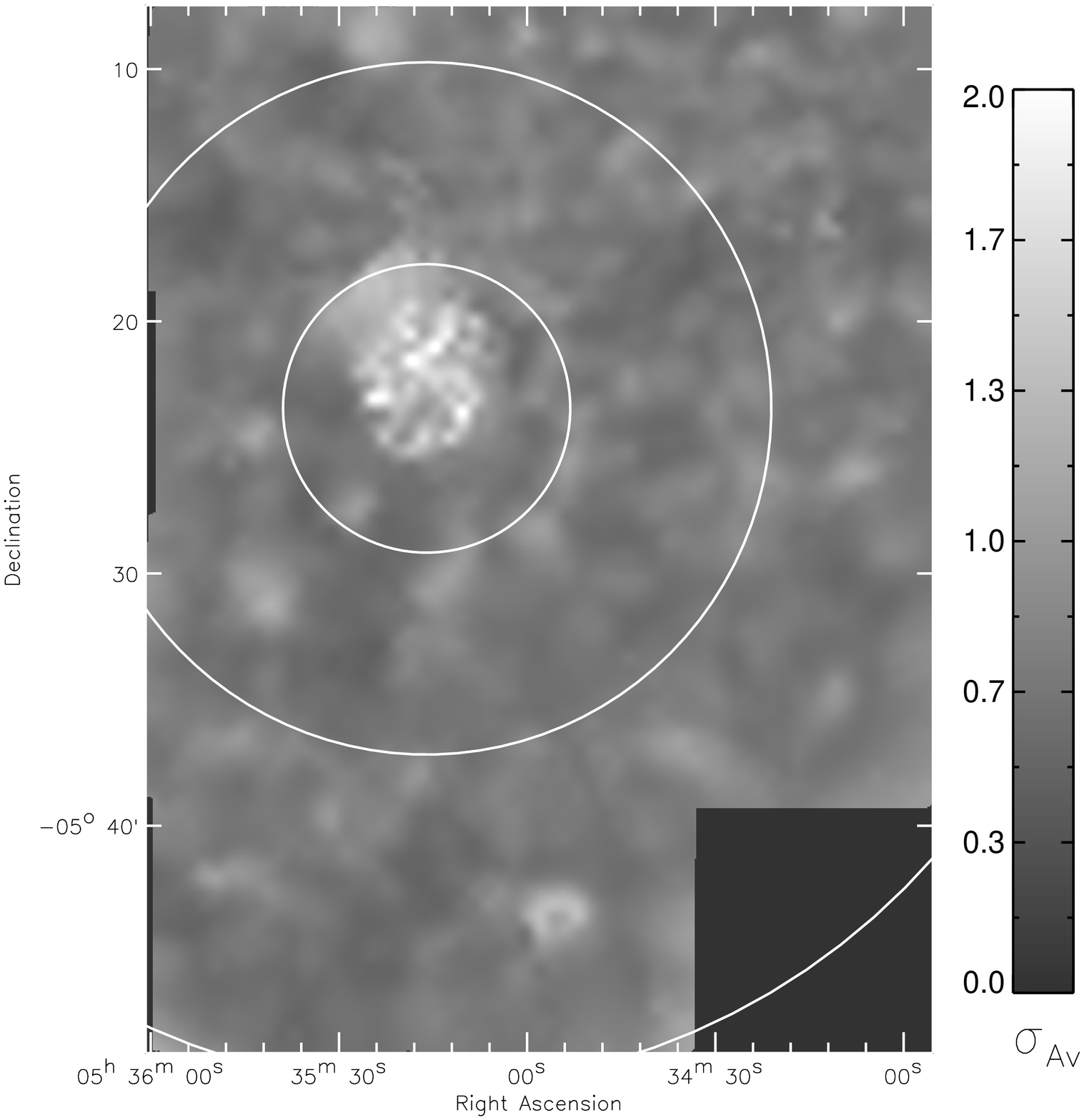}
 \caption{\textit{Left panel - }Extinction map of the OMC-1 derived as described in the text. The color bar shows the correspondence between the gray scale and visual extinctions. We also mark the boundaries of the three regions outlined by R10. \textit{Central panel - }ISPI \kb\ mosaic of the surveyed ONC. The overplotted contours represent the A$_V$=10 (yellow line) and the A$_V$=20 (red line) levels. \textit{Right panel - }The error map for our derived OMC-1 extinction map.
 The OMC-1 extinction map and the corresponding error map are also available in electronic form at the CDS
  .}\label{fig:buildavmap}
\end{figure*}

It turns out that, particularly within a few arcminutes from the Trapezium cluster, the number of predicted background stars is lower than the number of observed ones. The excess sources are, most probably, ONC members wrongly identified as background stars, having photometry matching our background selection criteria. To statistically reject these extra sources, for each group of 20 stars we discard the background star with the lowest extinction $A_{V,min}$, assuming it is a cluster member, and replace it with a new, further background star with  $A_V>A_{V,min}$, if existing (we are still constrained by the 1,000 pixels maximum distance). By decreasing the density of background sources, we obtain an extinction map which matches both the typical reddening and source density of the galactic component, averaged over 5\arcmin, or less. The result is shown in Fig.~\ref{fig:buildavmap}
, where we have interpolated the discrete gridding of the map using the IDL \texttt{GRIDDATA.pro} routine. The angular resolution ranges between 1.5\arcmin\ and 5\arcmin, depending on the density of background sources (Fig.~\ref{fig:densities}). In correspondence of the highest extinction values the map reaches its worst resolution, as expected.

The error map shown in Fig.~\ref{fig:buildavmap} shows a typical uncertainty $A_V\lesssim$1. The regularity in this map is due to the fact that for each grid point, apart the Trapezium region, the algorithm collects the requested 20 stars within a distance shorter than the limit of 5\arcmin. It follows that the number of stars in the analyzed subsample (corresponding to the given grid point) does not depend on the stellar density, but on the combination of the density and the angular resolution of the extinction map.

\subsection{Biases in the selection of background stars}
\subsubsection{Foreground contamination}
In deriving our extinction map we have assumed that the foreground contamination is closely modeled by the Besan\c{c}on synthetic galactic population. It is otherwise well established that the ONC (also known as Orion subgroup Id) is neighbored by three somewhat older subgroups of stars \citep[Ia, Ib, and Ic,][]{Brown}, which are located at distances ranging from $\sim$360 to 400 pc and have ages from $\sim$2 to 11.5 Myr. The most likely subgroup from which we would see contamination in our data is subgroup Ic, which is located along the same line of sight of the ONC and is thought to be as young as 2 Myr old \citep{Brown}. Thus, we expect little differences between the ONC and subgroup Ic isochrones, the main source of scatter being the smaller distance modulus, providing photometry up to 0.3 magnitudes brighter.

Considering that subgroup Ic is, at most, as reddened as the ONC by interstellar extinction, we expect that the isochrone shown in Fig.\ \ref{fig:contcm} also applies to the CMD of subgroup Ic. Thus, we still expect to find in each neighborhood the two main populations discussed above, the least reddened containing both ONC and subgroup Ic members, together with foreground field stars. Therefore, the presence of spurious sources belonging to the Ic subgroup should not affect our computation of the OMC-1 extinction map.

\subsubsection{Contamination by extragalactic sources}
In deriving the OMC-1 extinction map, we model the background sample with the reddened galactic model but neglect any contamination by extragalactic sources. We now address the reliability of this assumption.

To compute the extragalactic counts model reddened by the OMC-1, we use a deep NIR catalog of galaxies based on SOFI observations in an area of 340 ${\rm arcmin}^2$, centered on the coordinates RA=$3^h 32^{\prime} 28^{\prime\prime}$ and DEC=$-27^{\circ} 48^{\prime} 27^{\prime\prime}$ (kindly provided by T. Dahlen,\ \textit{private communication}). At these coordinates, the interstellar galactic extinction is particularly low (SFD98 extinction map provides $A_V\sim 0.02$), entirely negligible compared to the extinction provided by the OMC-1. Thus, assuming that the distribution of galaxies beyond the OMC-1 is well reproduced by the SOFI catalog, we randomly and uniformly spread the sample of galaxies over the field. Then, assuming that the main source of extinction is provided by the OMC-1, we add the proper amount of extinction based on our map, converting from visible to the infrared wavelengths via the extinction curve given by \citet{Cardelli00}:
\begin{eqnarray}
 A_{J}=0.288A_V;\ A_{H}=0.182A_V;\ A_{K_S}=0.118A_V\label{eq:cardelli}
\end{eqnarray}

We repeat this procedure 1,000 times, and we average the sample of 1,000 extragalactic CMDs, in order to minimize statistical fluctuations. We find that the reddened extragalactic population accounts for just $\sim$1\% of the observed background sample in Fig.\ \ref{fig:contcm} down to \hb$\sim$17. This is roughly the same fraction of galaxies found by \citet{Ricci2008} in their analysis of the deep ACS images of the Orion Nebula, and confirms our hypothesis of negligible extragalactic contamination. We also remark that we removed all the galaxies resolved by ACS from the analyzed sample, thus further minimizing the incidence of extragalactic sources.

\section{The Orion Nebula extinction map}\label{sec:foreground}

It is known that most of the extinction toward the Trapezium cluster arises at the interface between the M42 HII region and the neutral diffuse matter \citep[see][and references therein]{ODell2009}.  As indicated by \citet{odell2000}, internal extinction may also play a role. Aiming at mapping  the extinction provided by the diffuse matter in front of the OMC-1 (the \lq\lq Orion Nebula extinction\rq\rq\ hereafter), we focus now our attention on the sample of stars which are candidate members of the ONC. All these sources have been detected in the \jb, \hb\ and\kb-bands.

\begin{figure*}
 \centering
 \includegraphics[width=.9\linewidth]{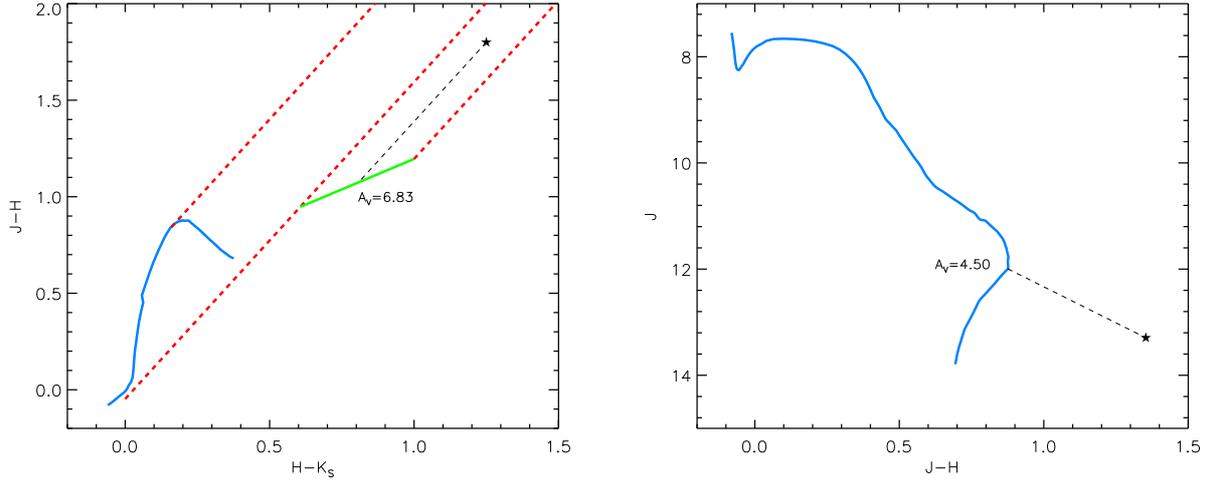}
 \caption{\textit{Left panel - }Example of the dereddening algorithm, coded with the black dashed line, for a star with colors compatible with the reddened CTTSs locus provided by \citet{Meyer97} (the green solid line). The blue solid line is the 2 Myr old isochrone suitable for the ONC. The red dashed lines are parallel to the reddening direction and bracket the regions occupied by the previous loci reddened by any A$_V$. \textit{Right panel - }Example of the reddening algorithm for a star with NIR colors compatible with reddened photospheric colors. The blue line represents the isochrone in the (\jb,\jb-\hb) diagram.}\label{fig:dereddening}
\end{figure*}

We face now a different set of problems. Whereas one can safely assume that the 2~Myr isochrone provides a relatively accurate locus for the dereddened stellar photospheres, one has to consider the spurious presence of circumstellar matter, disks in particular, which may affect the brightness (and to a lesser extent the IR colors) of the sources. Depending on the typical disk flaring angles, which are enhanced by the UV radiation within the HII region \citep{Robberto02} the sample of edge-on disks may be relevant. We shall assume that they still represent a small contamination, since they tend to move the sources beyond our \kb=15 limit for cluster membership.  Probably more important is the fact, also found by R10, that $\sim$11\% of the point-like sources in the ISPI catalog do not show NIR colors compatible with reddened photospheric colors. Their position in the (\jb-\hb,\hb-\kb) diagram is compatible with the reddened CTTSs locus introduced by \citet{Meyer97} (Fig.\ \ref{fig:dereddening}), i.e.\ their photometry is indicative of strong excess emission from circumstellar disks. Following \citet{Mey96}, we split our candidate members sample in two subgroups. If the observed colors are compatible with reddened CTTSs colors, then we deredden the observed colors taking the CTTSs locus as reference. Otherwise, in order to minimize the effects of eventual NIR excesses from circumstellar disks, we deredden our sources in the (\jb,\jb-\hb) diagram taking the 2 Myr old isochrone shown by R10 as the reference locus. In both cases, our dereddening algorithm moves the observed photometry to the reference locus (either the CTTSs locus or the isochrone) along the reddening direction, computed using the \citet{Cardelli00} reddening law. Figure \ref{fig:dereddening} graphically shows our dereddening algorithm.

\begin{figure*}
 \includegraphics[width=.35\linewidth]{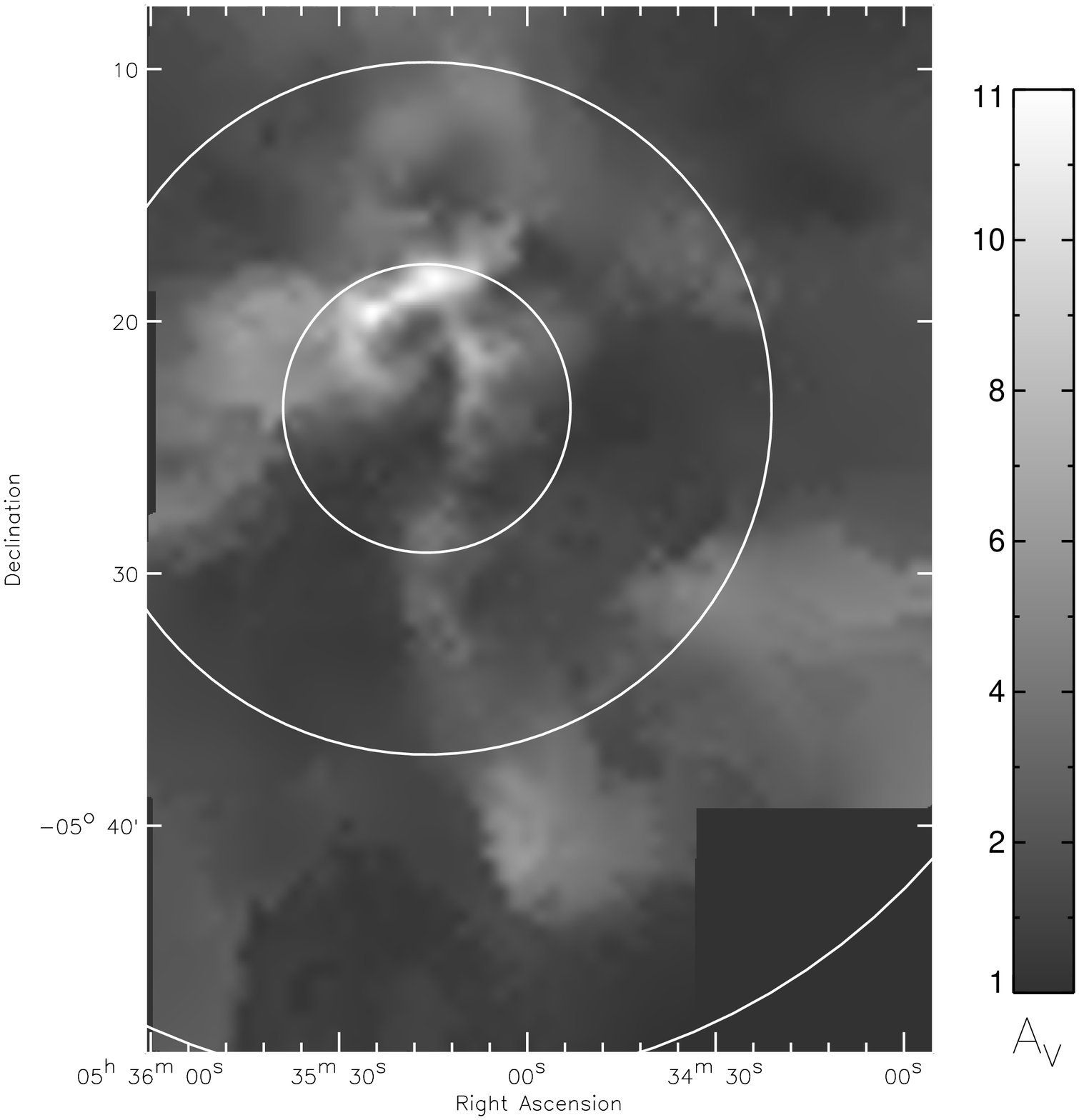}
 \includegraphics[width=.29\linewidth]{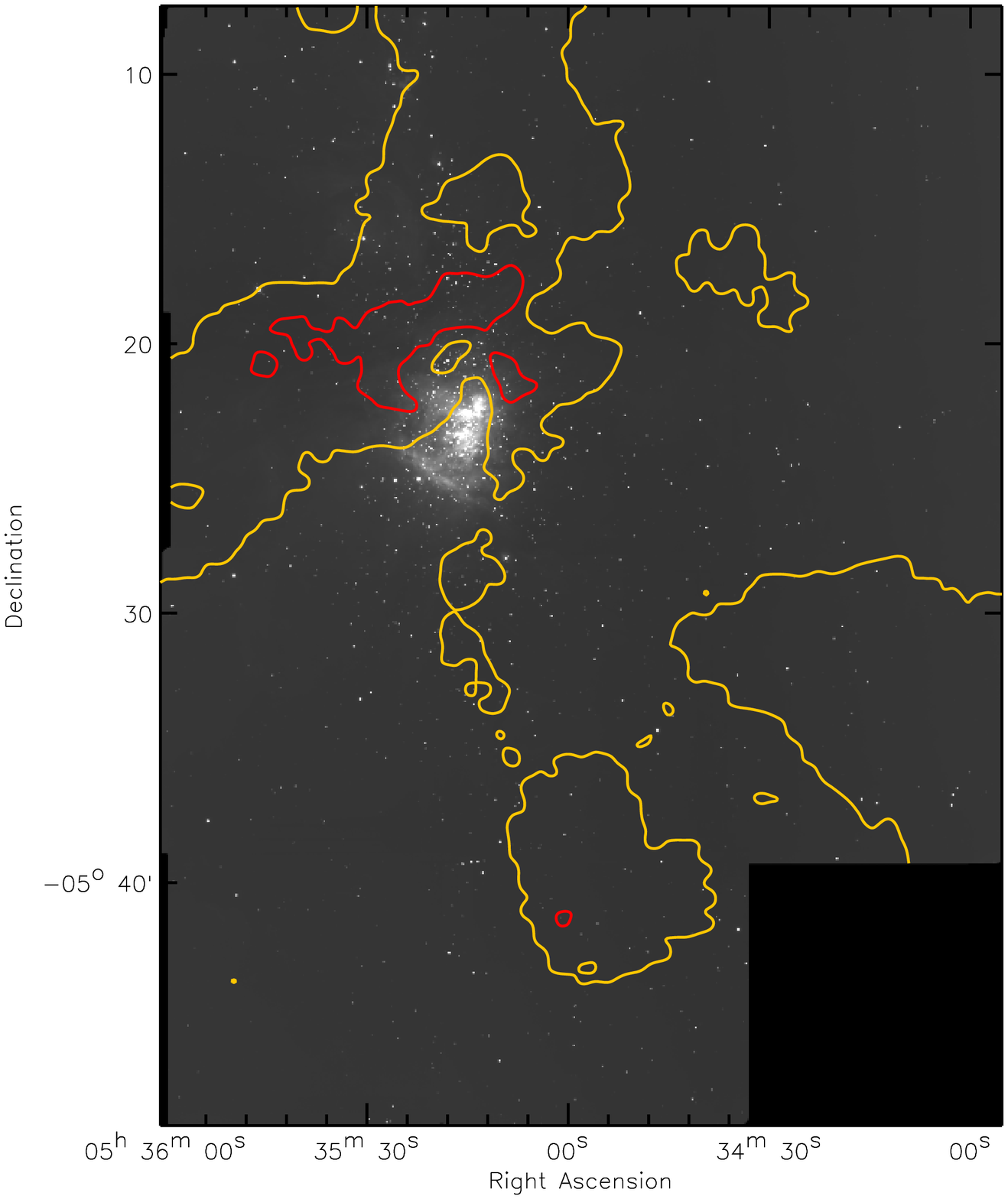}
 \includegraphics[width=.35\linewidth]{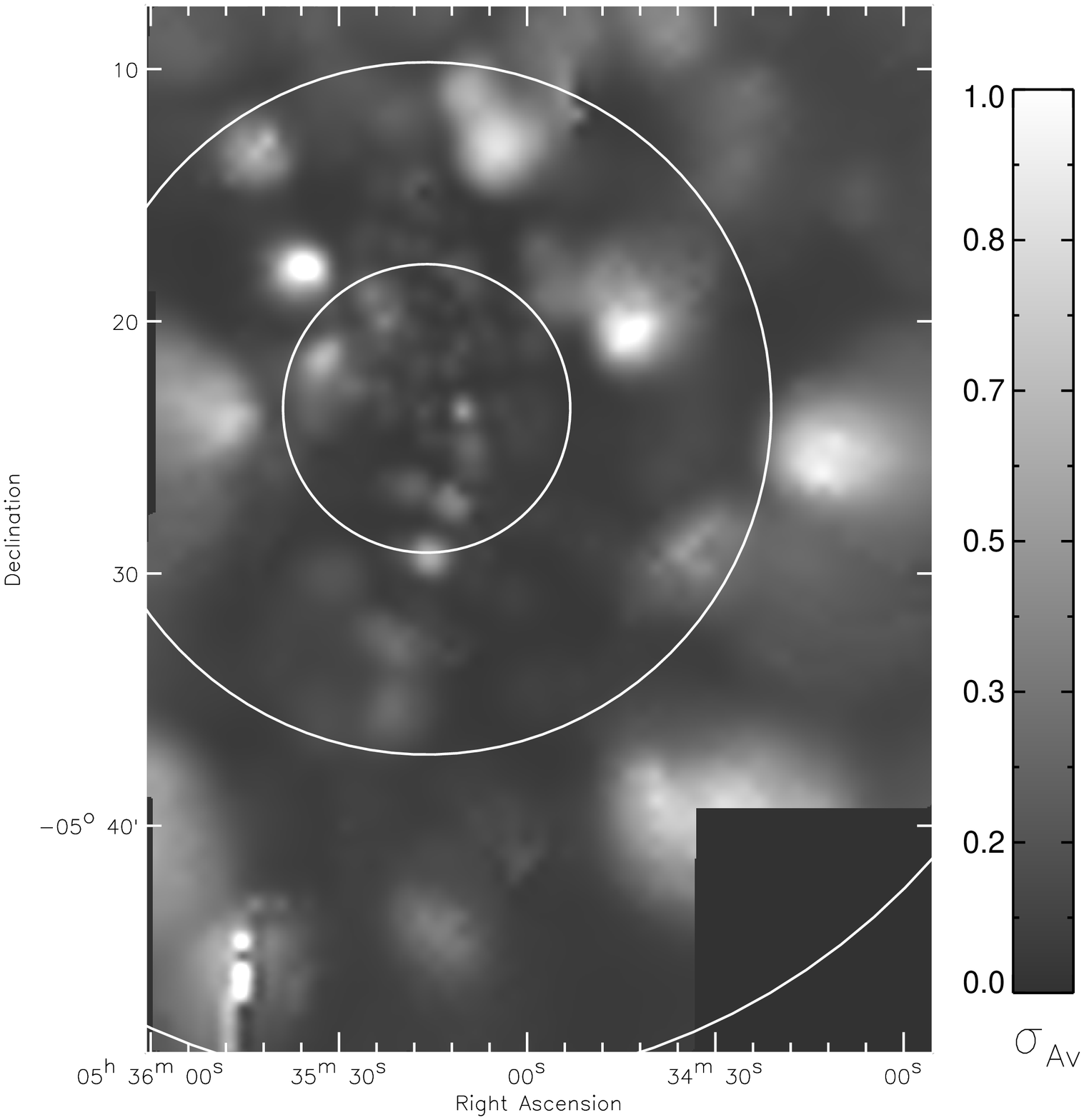} \centering
 \caption{\textit{Left panel - }Map of the average extinction affecting the ONC candidate members. The color bar shows the correspondence between the gray scale and visual extinctions. We also mark the boundaries of the three regions outlined by R10. \textit{Central panel - }ISPI \kb\ mosaic of the surveyed ONC. The overplotted contours represent the A$_V$=3 (yellow line) and the A$_V$=6 (red line) levels. \textit{Right panel - }Uncertainty on our derived ON extinction map.
 The Orion Nebula extinction map and the corresponding error map are also available in electronic form at the CDS
 .}\label{fig:veilmap}
\end{figure*}

The set of A$_V$ estimates is then locally averaged following the same strategy  described in Sect.~\ref{sec:computation}. In this case, we adopt a maximum angular resolution of 1,500 pixels ($\sim$7.5\arcmin) to allow our algorithm to find a good number of cluster members in  the outskirts of the Nebula, where the surface density of stars is generally low (see Fig.~\ref{fig:densities}). The resulting extinction map, with the associated error map, are shown in Fig.~\ref{fig:veilmap}
.

We find that in the direction of the Trapezium cluster the ON extinction is generally $A_V\lesssim$3, with a clear increase ($A_V\gtrsim$6) in correspondence of the dark ridge along the north-east edge of the HII region, which is therefore a foreground structure. The ridge is part of a larger bow-shaped feature extending from the east to the north of the ON. These findings are fully consistent with the map derived by \citet{odell2000} combining radio and optical measurements over an area 360\arcsec$\times$425\arcsec\ wide around $\theta^1$Ori-C.


\section{Discussion}\label{sec:discussion}

\subsection{Structure of the OMC-1 beyond the Orion Nebula Cluster}\label{sec:structure}

Alongside with the previous studies illustrated in Sect.~\ref{sec:overview}, our map (Fig.~\ref{fig:buildavmap}) shows that the OMC-1 surface density generally increases towards the Trapezium. In particular, we find that the Trapezium cluster is located in front of a high-extinction region ($A_V\gtrsim30$) extending $\sim$10\arcmin\ to the north of $\theta^1$Ori-C. This region is delimited to the south-east side by a sharp edge: the optical thickness of the OMC-1 decreases steeply by $\sim$20 magnitudes in a few arcminutes. This edge corresponds to the Orion Bar, the bright feature directly discernible in the ISPI image (right panel in Fig.~\ref{fig:buildavmap}). The steep drop of extinction can be confirmed even by a direct inspection of the images. It is quite evident that the density of faint and red stars suddenly increases to the south of the Orion Bar. A similar extinction drop is found also to the north-eastern edge of the extinction peak, in correspondence of the dark structure, known as the Dark Bay or Fish Mouth, seen in absorption in the optical images of the ON. Elsewhere, the extinction smoothly decreases down to $A_V\gtrsim4$ with increasing distance from the Trapezium cluster.

On larger scales, the OMC-1 extinction map shows a north-south pattern. The elongated extinction ridge is distributed over the full extent of the survey, reaching the OMC-2/3 star forming region to the north of the ONC \citep{Peterson2008}. The extinction map thus follows the dense filament traced by the molecular column density data of \citet{gold97}.

\begin{figure}
 \centering
 \includegraphics[width=.9\linewidth,viewport=0 0 425 340,clip]{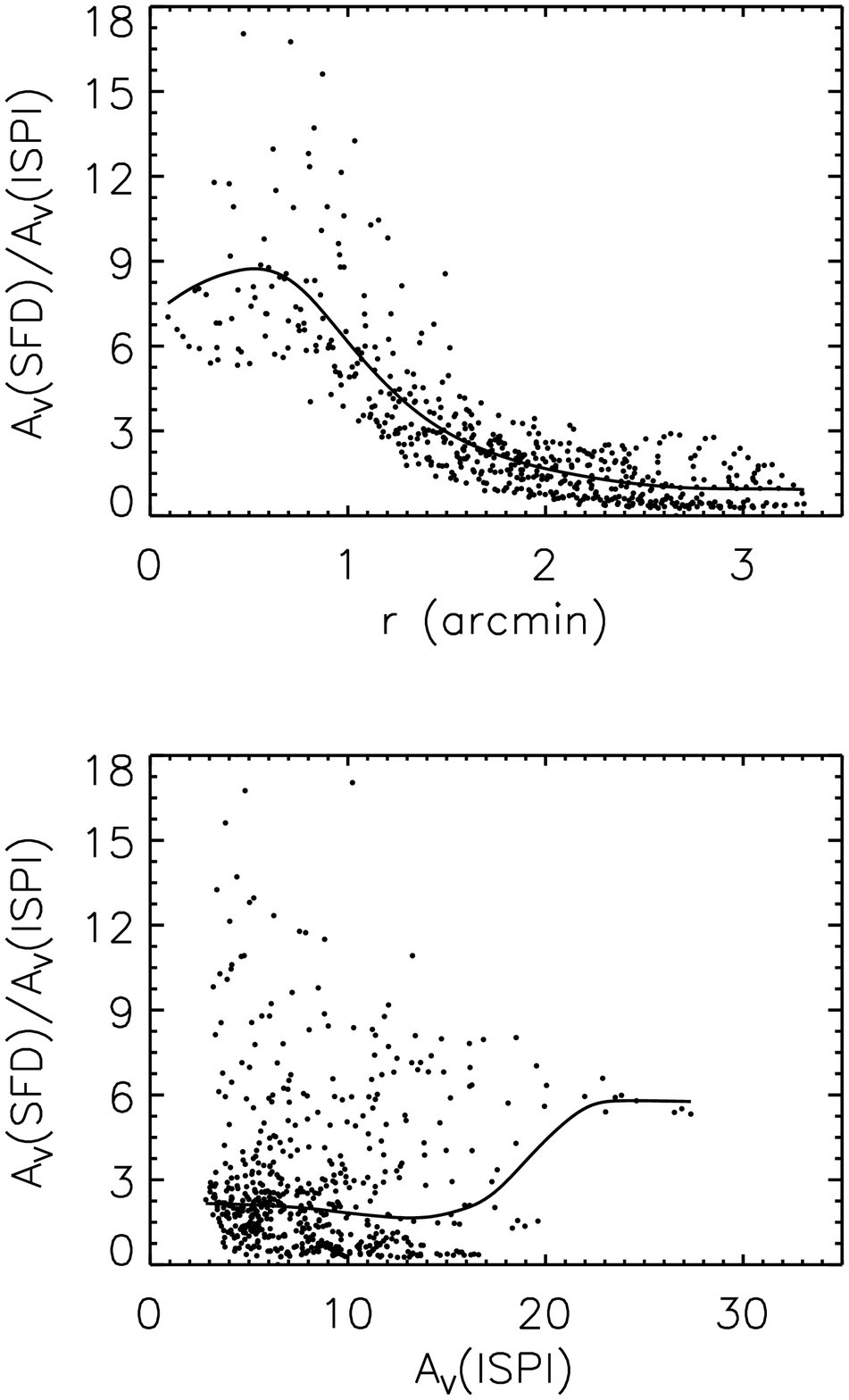}
 \caption{Diagram of the ratio $r_{A_{V}}$ as a function of $A_V(ISPI)$ computed over our extinction map. The solid line is a binned sample median, aimed at stressing the overall trend of the ratio. This diagram shows that for low extinctions our measurements and the SFD98 extinction are comparable, and that the latter becomes up to 3 times larger for higher extinctions, as found by \citet{dobashi05}.}\label{fig:avmap_comparison}
\end{figure}

To compare our extinction map with the one of SFD98, we degrade our spatial resolution down to their value. The pixel-by-pixel ratio $r_{A_{V}}=A_V(SFD)/A_V(ISPI)$ is plotted in Fig.~\ref{fig:avmap_comparison} as a function of $A_V(ISPI)$. For low extinction values, the SFD98 values generally overestimate the extinction by a factor of ~1.5--2, consistently with what found by \citet{Arce99}. Moreover, consistently with \citet{dobashi05}, the SFD98 values are systematically larger than ours, the ratio increasing in the $3\lesssim r_{A_{V}}\lesssim5$ range with the $A_V(ISPI)$ extinction. This is indicative of the fact that SFD98 map is not accurate either in high extinction regions or in regions with high extinction gradients.

\subsection{Structure of the Orion Nebula}\label{sec:on_structure}

As proposed in the 3-D model of the ON by \citet{ODell2009}, the boundary between the HII region and the bulk of the neutral matter is made up by a geometrically thin and optically thick shell swept up by stellar winds. Given this model, we argue that the extinction in the Dark Bay direction and the broad feature around it are spatially correlated, the former being a clump of high column density gas inside the remnant of the neutral shell located between the observer and the Trapezium \citep[see Fig.~4 in][]{ODell2009}.

On larger scales, the ON is never thinner than $A_V\sim$2 and this is compatible with the presence of a foreground neutral veil \citep{ODell2009}. Furthermore, a north-south pattern is evident, similarly to what is found for the OMC-1 extinction map (Fig.\ \ref{fig:buildavmap}). This pattern roughly follows the north-south elongation of the ONC observed by \citet{HillenbrandHartmann1998}, indicating that the cluster is still partially embedded in the OMC-1. As pointed out by \citet{HartmannBurkert2007}, this suggest that it might be better to view the ONC as a dense, moderately elongated cluster embedded in a larger scale filamentary stellar distribution rather than a system having a smoothly varying ellipticity as a function of radius. As indicated by \citet{HillenbrandHartmann1998}, the close association between the distribution of young stars and the elongated diffuse structures suggests that no dynamical relaxation has taken place so far, i.e.\ the stellar population has not dynamically adjusted from its initial state. 


\section{Conclusions}\label{sec:conclusion}

In this paper we have used the most recent wide-field NIR photometric catalog to discriminate between two main families of stellar sources in the ON region: 1) ONC members, distributed over an area $\sim$30\arcmin$\times$40\arcmin\ roughly centered on $\Theta^1$Ori-C; 2) background stars. The two samples have been statistically analyzed to derive the extinction maps of the OMC-1 (Sect.\ \ref{sec:computation}) and the foreground ON (Sect.\ \ref{sec:foreground}), with angular resolution $<$5$\arcmin$ and $<$7.5$\arcmin$ respectively.

Our results show that the OMC-1 generally accounts for the largest amount of extinction, typically $A_V\gtrsim$6, steeply rising up to $A_V\gtrsim$30 in the direction of the Trapezium cluster and dramatically dropping south of the Orion Bar.

The extinction toward the ON is lower, $A_V\lesssim$3, with a peak value $A_V\sim$6 in the direction of the Dark Bay feature. Our findings agree with \citet{Hill97}, \citet{Mue02} and \citet{DaRio10} who derived that the ONC members are typically extincted by $A_V\lesssim$3.

Together with \citet{Arce99} and \citet{dobashi05}, we find that the OMC-1 extinction map proposed by SFD98 is overestimated by a factor of$\sim$3-5, especially in the optically thickest regions. Together with the comparability between our ON extinction map and the results of \citet{odell2000}, this supports both the robustness of our statistical approach and the general validity of our results over the full spatial extent of the Orion Nebula Cluster.

Our derived maps for the OMC-1 and the Orion Nebula are available in electronic form as FITS files at the CDS. 


\appendix

\section{Completeness}\label{sec:completeness}
To determine the completeness of the R10 photometric catalog, we use the artificial star experiment performed by R10, increasing the number of tests up to 10$^4$ per magnitude and per field. Our completeness estimation turns out to be a \lq\lq local\rq\rq\ procedure as it provides the completeness sensitivity at any given position and for any given magnitude, computed within a circle with a radius of $\sim$1$\arcmin$, i.e.\ lower than the typical size of the nebular structure present in the ISPI images. The number of tests allows us to estimate the 100\% completeness with an error of a few 1\%.

According to R10, we find that the completeness (and the sensitivity)\ of our survey decreases with decreasing distance from the Trapezium. This is due to the combined effects of increasing crowding and nebular brightness close to the inner cluster.
       
\begin{figure}
 \centering
 \includegraphics[width=.9\linewidth]{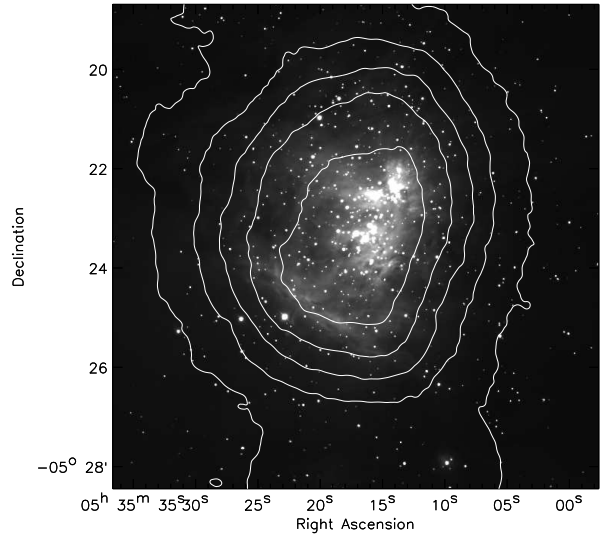}
 \caption{Completeness contours for \kb=17 in the neighborhood of the Trapezium stars. The contours represent the 10\%, 30\%, 50\%, 70\% and 90\% levels with increasing distance from the inner cluster respectively. This figure shows that the iso-sensitivity contours roughly follow the iso-brightness levels of the image, indicating that the completeness of the survey is strongly affected by the background nebular emission.}\label{fig:completeness}
\end{figure}

The results of our simulations are shown in Figure \ref{fig:completeness}, reporting the completeness trend for a \kb=17 star in the inner surveyed region. As we stated above, we find that the sensitivity of the survey gets drastically shallower in the very inner region, decreasing from $\lesssim$100\% down to $<$10\% in a few arcminutes.

\begin{acknowledgements}
We thank Prof.\ C.~R.~O'Dell for providing his useful comments on this paper.
\end{acknowledgements}


\end{document}